# Measurement of cabin pressure during a flight with a smartphone

Régis Barillé

Université d'Angers, CNRS, MOLTECH-ANJOU, SFRMATRIX, F-49000 Angers, France

**abstract**

An aircraft cabin is used as a laboratory for studying the atmospheric pressure during a flight. All the different steps of the flight – take-off, cruise altitude climbing and landing - are monitored with the use of the pressure sensor of a smartphone. Specific details of the atmospheric pressure during take-off and landing are given. Calculations are done for the aircraft's nose elevation angle during take-off with the relation between the optimized cabin pressure for the passenger's comfort and the aircraft altitude.

Air passenger traffic has now become a useful way for more and more people to travel. However, during a flight it is interesting for a curious passenger to wonder: what is the current cabin pressure inside the airplane? In particular, if passengers travel at high altitudes, what is the angle of elevation of the plane during take-off?, what is the cabin pressurization behavior of the aircraft during landing? In this goal a smartphone can be used for responding to these questions. The size reduction of electronic devices has increased the integration of a large range of sensors in the new generation of smartphones. Currently, the smartphone is an incredible instrument that has a wide range of sensors (accelerometers, gyroscopes, magnetometer, etc.) that can be used to study a large choice of physics topics from mechanics [1] to electromagnetism [2]. Different experiments have shown its use for teaching physics to students [3]. Smartphones open the possibility of designing and developing low-cost laboratories where expensive testing devices can be substituted by smartphones [4]. Recently for example, the characteristics of the inner layer of the atmosphere were analyzed using a smartphone mounted on a drone. In the study, the pressure was obtained using the smartphone's pressure sensor and compared to different models of the atmosphere [5]. Another study uses the pressure sensor, or barometer, present into usual smartphones to obtain vertical velocity data in several experiments: in the elevators of tall buildings, in pedestrian climbing stairways, and on drones [6]. An aircraft can be used as a physics laboratory to monitor the pressure inside the cabin during the different moments of a flight. A previous experiment was done with the use of a water-filled manometer to determine the drop in cabin pressure during the flight [7]. In the experiment, a sample of air inside a syringe is trapped. If the volume and temperature of the sample is kept constant, the pressure of the sample is also constant. The sample can then be used as a fixed pressure reference against changes in the cabin pressure and the variations can be determined. Here in the present study, the cabin pressure is monitored with the pressure sensor of a smartphone during the different steps of a flight (take-off - landing). The expected results can help to understand the relation between pressure/altitude and the pressurization/repressurization effect during a flight for the fuselage safety during take-off and landing.

The smartphone used in the experiment is a Samsung galaxy S10 with a MEMS pressure sensor (STMicroelectronics) with a pressure range of 26 kPa - 126 kPa and a resolution of $2.10^{-5}$ kPa (1 hPa = 100 kg/m.s$^2$). The device is able to detect absolute pressure with a pressure noise lower than a root mean square pressure of 1 Pa.

The data are monitored and acquired with the Phyphox application [8]. The acquired data are exported in a 'CSV' extension file for calculations.

The smartphone in this experiment is used to observe the behavior of the pressure inside the cabin of an aircraft that passengers experience during take-off and landing. In the experiment, the phone's pressure sensor is used to measure the pressure inside the cabin of a Boeing 737-type aircraft. The phone is laid flat on the tray table of the seat back.

Boeing 737 is a short-to medium-range, narrow-body aircraft. The Boeing 737 type air-conditioning and pressurization system of the cabin air is supplied by 'bleed air' taken from the engines. The continuous inflow of conditioned air pressurizes the cabin, with excess air vented back to the atmosphere through the outflow valve. The cabin pressure is obtained by controlling the flow of air which enters and leaves the cabin. The flow is managed by the pressurization controller.

Measurements begin when the aircraft is taxiing on the runway. During the take-off phase preparation (Fig. 1a), the atmospheric pressure starts from a pressure of 101.4 kPa. This pressure is the ambient aircraft cabin pressure considered in the calculation as the ground reference pressure. The value is adapted for the comfort of the passengers with a temperature of 18 °C. A pressure variation was observed during the take-off between the times 0 and 40 s. First a pressure decrease is measured followed by a sudden rise of the cabin pressure until 102.2 kPa. On a number of aircraft types the cabin pressure is increased during the take-off roll, ensuring there is a slight over-pressure when the aircraft rises up after take-off. This pressure increase prevents some sudden changes in the cabin pressure during the aircraft rotation when the cabin exhaust valve is exposed to a quick and large change of airflows. The pressure variation corresponds to a virtual altitude value below the ground. The measurement of the maximum value of 102.2 kPa corresponds to a differential pressure increase of 0.827 kPa (0.12 psi) with the initial ambient cabin pressure. The measured value is in agreement with the cabin differential pressure given by the manufacturer [9].

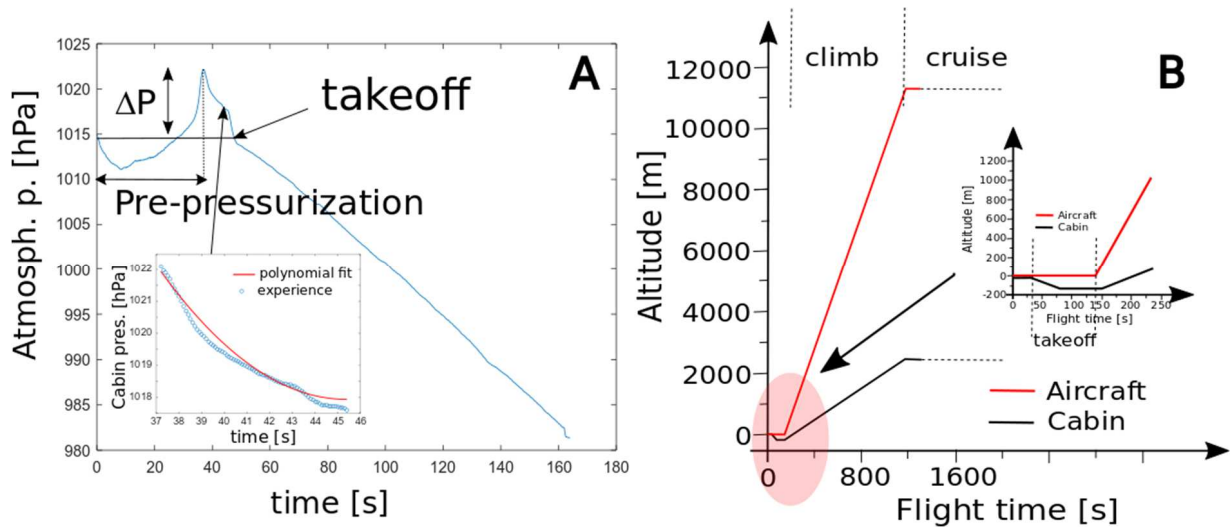

Fig. 1: Variation in the cabin pressure of an airplane during take-off and climbing a) experimental results, b) aircraft pressurization dynamics as a function of time

After the time corresponding to the maximum pressure, the aircraft accelerates and the cabin pressure decreases non linearly. The time evolution of the pressure can be fitted with a function: $p = a_1 t^n + a_2 t + a_3$ with $n \approx 2$. The quadratic evolution of the pressure shows that the aircraft speed v(t) (Δelevation/Δtime) is linear and the acceleration is a constant. This relationship is related to the Bernoulli's equation. At t = 45.8 s the aircraft reached its take-off safety speed and the aircraft nose is raised above the horizon. The time duration could be evaluated by evaluating the slope of the pressure variation. The nose raises from t = 45.8 s to t = 48.9 s (Fig. 1a). During this period the cabin pressure varies from 101.77 kPa to 101.38 kPa and corresponds to an altitude variation of a = 31 m. The altitude is calculated based on the calculation of the pressure variation with the elevation:

$$p(z) = p_0 e^{-\frac{Mg}{RT}z}$$

where p is the altitude pressure in pascals, z the altitude in meters, M = 28.966 $10^{-3}$ kg.mol$^{-1}$, g = 9.805 m.s$^{-2}$, R = 8.314510 J.mol$^{-1}$.K$^{-1}$, T = (273 + T0) K [10].
The aircraft speed in the configuration of take-off for a Boeing 737 has an average value of 268 km/h or 145 knots given by the manufacturers [11]. A simple calculation considering the distance traveled (b = 74.6 m/s * 3.1 s = 231 m) with this speed during the ascent Δt = 3.1 s gives the angle of plane nose elevation α:

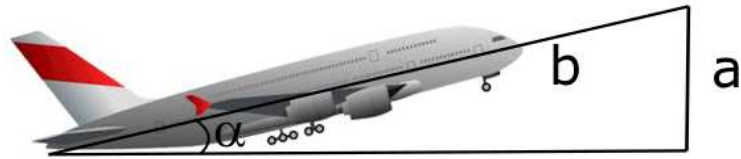

Fig. 2: Calculation of the nose elevation during take-off

$$\alpha = arcsin\left(\frac{a}{b}\right) * \frac{180}{pi} = arcsin\left(\frac{31}{231}\right) * \frac{180}{pi} = 7.75°$$

According to the manufacturer the angle of nose elevation (α) is set between 7° and 8° [11].

After take-off the aircraft noses up and climbs. The air pressure decreases as the aircraft gains altitude. The slope of the ΔP/Δt curve is used to calculate the aircraft's climb angle. For this calculation, we consider that the airplane has still the speed of 74.6 m/s (268 km/h or 145 knots) given by the manufacturer [12, 13] until it reaches the cruise speed. During a period of Δt = 11.5 s, the cabin pressure is changed from 101.38 kPa to 101.18 kPa. The aircraft is elevated with a high of 190 m or 629 ft and an ascending distance of 734 m is traveled. During the climb the cabin pressure is slowly decreased. The variation ΔP can be converted into a height variation Δh assuming the pressure difference between the cabin and outside according to the graph of both aircraft and cabin pressure evolution during altitude change (fig. 1b) as a function of time [14 - 15]. The cabin pressure is obtained by controlling the rates the air enters and leaves the cabin with the help of the pressurization system. The cabin's pressure altitude does not climb or descend as fast as the aircraft, and have its ceiling at 2439 m (8000 ft) to maintain oxygen pressure at a safe level to breath without a mask (fig. 1b). The pressurization system controls that on various phases of the flight the difference between the cabin pressure and the external pressure (differential pressure) does not overcome a pre-established value [16]. This difference is normally between 8 and 9 psi (55 kPa– 62 kPa). This value is less than the maximum one that the fuselage is able to support without suffering structural damages [17].

The maximal differential of pressurization of 0.861 kPa (0.125 psi) in the cabin for take-off must also be included [18]. In this case, 190 ± 5 m is obtained for the aircraft elevation. The angle is obtained by:

$$arcsin\left(\frac{a}{b}\right) * \frac{180}{pi} = arcsin\left(\frac{190}{734}\right) * \frac{180}{pi} \approx 15°$$

The value is in the range of the values given by the manufacturer Boeing for the V2 step take-off between 15° and 20° [19].

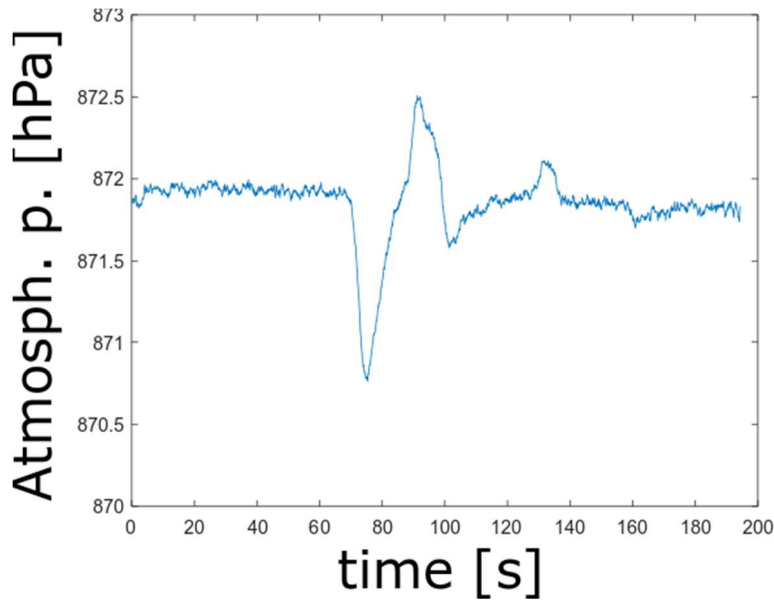

Fig. 3: Change in the cabin pressure of an aircraft during cruising flight.

After take-off, the aircraft maintains a speed to climb until it reaches the cruise altitude, then it maintains a constant speed of 969 km/h (Mach 0.785) at the aircraft's cruising altitude [20]. While a typical passenger jet may climb at speeds in excess of 280 km/h, the cabin pressure decreases at an equivalent rate of climb less than 152 m per minute. When the cabin pressure has reached an equivalent of about 2300 m ± 100 m (~8000 ft and 1 ft = 0.3 m) altitude (the aircraft is then higher by that time) the cabin pressure is kept constant.

The cabin pressure at the cruise altitude reaches a value of 87.2 kPa (Fig. 3). This cabin pressure can be converted in the equivalent cabin altitude of 2285 m or 7496 ft. The aircraft pressure altitude versus the cabin pressure graph gives the equivalent aircraft altitude (Fig. 1b). A graph is used giving the relationship between the environment pressure and the flight altitude of the aircraft or the cabin altitude versus the aircraft altitude. For example, an airplane at an altitude of 15,000 m and environmental pressure protect to 14.75 kPa (2.14 psi) results in a cabin pressure results of 2.14 + 8.9 = 11.04 psi (76.11 kPa) according to the typical flight pressurization profile (Federal Aviation Administration) [14]. The aircraft's cabin pressure will be equivalent to an altitude of 2433 m. A value of 10331.5 m (33 896 ft) is found in agreement with the usual altitude value for this type of aircraft and in agreement with the normal cruising altitude given by the manufacturer [21, 22]. Note that the cruise altitude is normally determined by winds, weather, and other factors. The optimal altitude is the altitude that gives the best fuel economy for a given configuration and gross weight.

During the flight sudden small variations of the cabin pressure can be measured with the smartphone suggesting that the aircraft has a non-constant or fluctuated altitude (Fig. 3). It could also be caused by the balance of air inflow and outflow when the cabin pressure regulator controls the opening and closing of the aircraft's outflow valve. A variation of the cabin pressure of ΔP = 0.17 kPa can be observed. The variation ΔP corresponds to a change in the altitude of 15.9 ± 0.2 m.

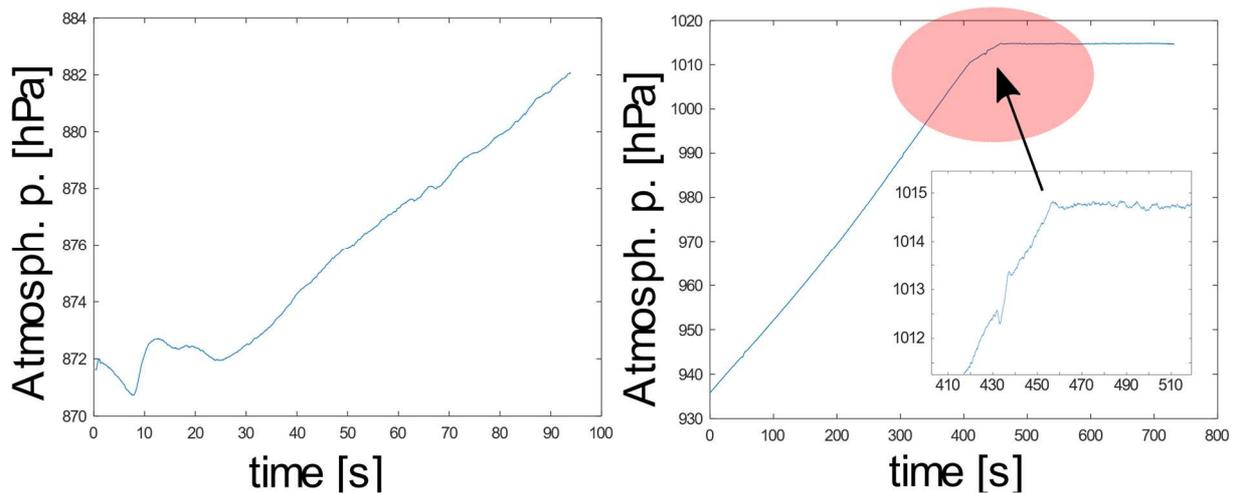

Fig. 4: Pressure inside the aircraft descent for landing to the airport.

As the plane begins its descent to land, the change in atmospheric pressure approaches the one of the ground and increases. The observation of the descent stage can be done by examining the pressure variation with the smartphone (Fig. 4). First an oscillating variation appears. The behavior of the response is nonlinear. The response gives the second order transition and corresponds to a second order impulse response with an underdamped system. The system's response is provided by the application of an air pulse to the air pressurization management. Then, after the oscillating transition, the cabin pressure is continuously increased [23].

During the landing period when the aircraft is 9 m (30 ft) above the runway, it flares up by raising the nose about 5 degrees and the aircraft lands on the runway.

When the wheels are placed on the track, the pressure varies abruptly. Note that the pressure when the aircraft is close to the ground is lower than the cabin pressure when the aircraft touches the ground (Fig. 5). The cabin pressure continues to descend towards the altitude of the arrival terrain minus the equivalent of 0.69 kPa (0.1 psi) to avoid jerks during landing. The differential pressure given by the pressurization controller can be calculated and is used for the same reason than take-off. It ensures there is a slight over pressure when the aircraft lands. The measurement gives a jump in the cabin pressure of 0.72 kPa and it corresponds to a value of 0.1 psi according to the manufacturer.

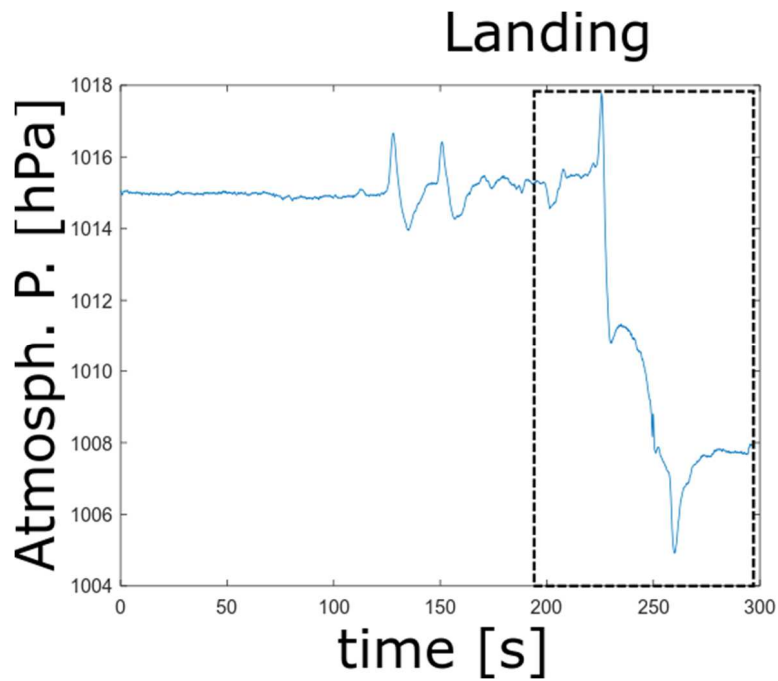

Fig. 5: Change in atmospheric pressure when the aircraft is landing.

In conclusion the aircraft cabin can be used as a laboratory for experiments about atmospheric pressure. All the different stages during a flight can be monitored with the smartphone's barometer with a good resolution. The smartphone with its sensors and applications is able to follow the take-off and landing and gives details about the steps involved in the management of the cabin pressure for the comfort of the passengers. A last point partially mentioned here is the pulse response of the cabin pressure controller which can be explained in a further study.

The followed pressure variations inside an aircraft cabin can be a good introduction to the management of pressure in systems including submarine or scuba diving. The experiments can help students to consider the adaptation in a different environment where the pressure is different. Experiments on how to introduce this content into the introductory physics classes on thermodynamics can also be used with a vacuum storage container for food or with a tank for rainwater harvesting with the smartphone as a pressure sensor [24].